\documentstyle[aps,prb]{revtex}

\begin{document}
\draft

\title{The branch process of Cosmic strings}
\author{Ying Jiang\thanks{Corressponding author; E-mail: yjiang@itp.ac.cn}}
\address{CCAST (World Laboratory), Box 8730, Beijing 100080, P. R. China}
\address{Institute of Theoretical Physics, Academia Sinica, P. O. Box 2735,
Beijing 100080, P.R. China\thanks{mailing address}}
\author{Yishi Duan\thanks{E-mail: ysduan@lzu.edu.cn}}
\address{Institute of Theoretical Physics, Lanzhou University, Lanzhou
730000, P. R. China}
\maketitle

\begin{abstract}
In the light of $\phi$--mapping method and the topological tensor current theory,
the topological structure and the topological quantization of topological defects
are obtained under the condition that Jacobian $J(\phi/v)\neq0$. When $J(\phi/v)=0$,
it is shown that there exists the crucial case of branch process. Based on the implicit
function theorem and the Taylor expansion, the generation, annihilation and bifurcation
of the linear defects are detailed in the neighborhoods of the limit points and
bifurcation points of $\phi$--mapping, respectively.
\end{abstract}


\section{Introduction}

There have been rapid and exciting developments over the last decades on the
interface between paticle physics and cosmology\cite{rovelli1}. Particle physicists pursing
the goal of unification would like to test their theories at energy scales
far beyond those available now or in the future in terrestrial accelerators.
An obvious place to look is to the very early Universe, where conditions of
extreme temperature and density obtained. Meanwhile, cosmologists have
sought to understand feature of the Universe currently observed by tracing
their history back to that very early period. An exciting outcome of the
interplay between particle physics and cosmology is the cosmic string theory%
\cite{hindmarsh1,vilenkin1,vilenkin2}. It is strongly believed to solve the
short-distance problems of quantum gravity at the Plank scale by providing a
fundamental length $l_{str}=\sqrt{\hbar c/T}$, where $T$ is the string
tension, and provides a bridge between the physics of the very small and the
very large. The research in the topic of cosmic strings can help to explain
some of the largest--scale structure seen in the Universe today.

Past researches have mostly focused on the dynamical properties of the cosmic
strings\cite{3,4,5}, and all of them are based on some particular models.
In our previous work\cite{jiang}, by making use of the $\phi$--mapping topological
current theory\cite{duan3}, which play an important role in discussing the topological invariant
and structure of the physical system\cite{duan3,duan2,duan4,duan6,xu}, we have studied
the topological structure and topological quantization of cosmic strings without any
concrete model. In this paper, based on our previous work, 
we will study the generating, annihilating, colliding, splitting and merging of
cosmic strings in topology viewpoint, and give the branch conditions of cosmic
strings without any concrete model.

This paper is organized as follows: In section 2, a brief review of the
topological structure and the topological quantization of cosmic strings 
is given. In section
3, by virtue of the implicit function theorem, the creation and annihilation
of cosmic strings at the limit points is discussed. The bifurcation behaviour of strings is
detailed in the neighborhood of bifurcation point in section 4.

\section{Topological structure and topological quantization of cosmic strings}

Cosmic strings are linear defects\cite{mazur1} in four dimensional space-time
$X$, analogous to those
topological defects found in some condensed matter systems such as vortex
lines in liquid helium, flux tubes in type-II superconductors or
disclination lines in liquid crystal, and they are closely related to the
torsion tensor of the Riemann--Cartan manifold \cite
{duan2,desabbata1,hammond2,letelier1}. 

In vierbein theory, the torsion tensor is expressed by 
\begin{equation}
T_{\mu \nu }^A=D_\mu e_\nu ^A-D_\nu e_\mu ^A,\;\;\;\mu ,\nu ,A=1,2,3,4
\label{2.t}
\end{equation}
where $e_\mu ^A$ is the vierbein field. As in our previous work\cite{jiang},
by analogy with the 't Hooft's viewpoint\cite{thooft1}, to establish a
physical observable theory of space--time defect, we must first define a
gauge invariant antisymmetrical 2--order tensor from torsion tensor with
respect of a unit vector field $N^A(x)$ as follows 
\begin{equation}
T_{\mu \nu }=T_{\mu \nu }^AN^A+e_\nu ^AD_\mu N^A-e_\mu ^AD_\nu N^A=\partial _\mu A_\nu -\partial _\nu A_\mu  \label{2.00}
\end{equation}
where $A_\mu =e_\mu ^AN^A$ is a kind of $U(1)$ gauge potential. This shows
that the antisymmetrical tensor $T_{\mu \nu }$ expressed in terms of $A_\mu $
is just the $U(1)$ like gauge field strength (i.e. the curvature on $U(1)$
principle bundle with base manifold $X$ ), which is invariant for the $U(1)$%
--like gauge transformation 
\begin{equation}
A_\mu ^{\prime }(x)=A_\mu (x)+\partial _\mu \Lambda (x)  \label{2.cc}
\end{equation}
where $\Lambda (x)$ is an arbitrary function.

In order to study the
string theory, we should extend the traditional concept of topological currents\cite{duan3}
which have been used to study the topological properties of point like defects\cite{xu,hong}
, and introduce a topological
tensor current of second order from torsion.
From the above discussions, we can define a topological tensor current
$j^{\mu \nu }$ as the dual tensor of $%
T_{\lambda \rho }$ as follow 
\begin{equation}  \label{2.12}
j^{\mu \nu }=\frac 12\frac 1{\sqrt{g_x}}\epsilon ^{\mu \nu \lambda \rho
}T_{\lambda \rho } 
=\frac 12\frac 1{\sqrt{g_x}}\epsilon ^{\mu \nu \lambda \rho }(\partial
_\lambda A_\rho -\partial _\rho A_\lambda ).
\end{equation}

Very commonly, topological property of a physical system is much more
important and worth investigating mediculously. It is our conviction that,
in order to get a topological result, one should input the topological
information from the beginning. Two useful tools---$\phi $-mapping method
and composed gauge potential theory\cite{duan3,duan4}--- just do the work. As mentioned in our
previous works\cite{jiang,duan2}, the decomposation of $A_\mu (x)$ can be expressed by
\begin{equation}
A_\mu (x)=\frac{L_p}{2\pi }\epsilon _{ab}n^a(x)\partial _\mu n^b(x); \;\;\;\;n^a(x)=\frac{\phi ^a(x)}{||\phi (x)||},
\label{2.star}
\end{equation}
where $\phi^a(x)$ is the order parameter field of cosmic strings, and $L_p=\sqrt{\hbar G/c^3}$ is the Planck length introduced to make the
both sides of the formula with the same dimension\cite{duan2}.
With the decomposition of $A_\mu $ in (\ref{2.star}), $j^{\mu \nu }$ can be
expressed in terms of $n^a$ by 
\begin{equation}  \label{2.13}
j^{\mu \nu }=\frac{L_p}{2\pi }\frac 1{\sqrt{g_x}}\epsilon ^{\mu \nu \lambda
\rho }\epsilon _{ab}\partial _\lambda n^a\partial _\rho n^b,
\end{equation}
which shows that $j^{\mu \nu }$ is just an antisymmetric and identically
conserved 2-order topological tensor current. Because of
the topological property of $n^a$, we input the topological information
successfully. Obviously, $n^a(x)n^a(x)=1,$ and $n^a(x)$ is a section of the sphere bundle $%
S\left( X\right) $\cite{duan3}. The zero points of $\phi ^a(x)$ are just the
singular points of $n^a(x).$                                 

By making use of the expression of $n^a$ in (\ref{2.star}) and the Laplacian
relation in $\phi$-space
\[
\partial _a\partial _a\ln \Vert \phi \Vert =2\pi \delta (\vec \phi
),\;\;\;\;\;\;\;\;\partial _a=\frac \partial {\partial \phi ^a}, 
\]
the topological tensor current $j^{\mu \nu}$ can be rewritten in a 
compact form
\begin{equation}  \label{2.16}
j^{\mu \nu }=\frac 1{\sqrt{g_x}}L_p\delta (\vec \phi )J^{\mu \nu }(\frac
\phi x),
\end{equation}
where $J^{\mu \nu}(\frac \phi x)$ is the general Jacobian determinants  
\begin{equation}  \label{2.15}
\epsilon ^{ab}J^{\mu \nu }(\frac \phi x)=\epsilon ^{\mu \nu \lambda \rho
}\partial _\lambda \phi ^a\partial _\rho \phi ^b.
\end{equation}
It is obvious that $j^{\mu \nu }$ is non-zero only when $\vec \phi =0.$

Suppose that for the system of equations 
\[
\phi ^1(x)=0,\,\,\,\,\phi ^2(x)=0, 
\]
there are $l$ different solutions, when the solutions are regular solutions
of $\phi $ at which the rank of the Jacobian matrix $[\partial _\mu \phi ^a]$
is 2, the solutions of $\vec \phi (x)=0$ can be expressed parameterizedly by 
\begin{equation}  \label{2.5}
x^\mu =z_i^\mu (u^1,u^2),\,\,\,i=1,\cdot \cdot \cdot l,
\end{equation}
where the subscript $i$ represents the $i$-th solution and the parameters $%
u^I(I=1,2)$ span a 2-dimensional submanifold with the metric tensor $%
g_{IJ}=g_{\mu \nu }\frac{\partial x^\mu }{\partial u^I}\frac{\partial x^\nu 
}{\partial u^J}$ which is called the $i$-th singular submanifold $N_i$ in $X.$
For each $N_i$, we can define a normal submanifold $M_i$ in $X$ which is
spanned by the parameters $v^A(A=1,2)$ with the metric tensor $g_{AB}=g_{\mu
\nu }\frac{\partial x^\mu }{\partial v^A}\frac{\partial x^\nu }{\partial v^B}
$, and the intersection point of $M_i$ and $N_i$ is denoted by $p_i$. By
virtue of the implicit function theorem, at the regular point $p_i$, it
should be hold true that the Jacobian matrix $J(\frac \phi v)$ satisfies 
\begin{equation}
\label{2.dad}
J(\frac \phi v)=\frac{D(\phi ^1,\phi ^2)}{D(v^1,v^2)}\neq 0.
\end{equation}

More deeper calculation can lead to the total expansion of the string current
\begin{equation}
j^{\mu \nu }=\frac{L_p}{\sqrt{g_x}}\sum_{i=1}^l\frac{\beta _i\eta _i\sqrt{g_v%
}}{J(\frac \phi v)|_{p_i}}\delta (N_i)J^{\mu \nu }(\frac \phi x),
\label{2.000}
\end{equation}
or in terms of parameters $y^A=(v^1,v^2,u^1,u^2)$%
\begin{equation}
j^{AB}=\frac{L_p}{\sqrt{g_y}}\sum_{i=1}^l\frac{\beta _i\eta _i\sqrt{g_v}}{%
J(\frac \phi v)|_{p_i}}\delta (N_i)J^{AB}(\frac \phi y).
\end{equation}
where $\beta _i$ is a positive integer called the Hopf index\cite
{milnor1} of $\phi $-mapping on $M_i$ and $\eta _i=signJ(\frac \phi v)_{p_i}=\pm 1$ is the Brouwer degree\cite
{milnor1} of $\phi $-mapping. $\delta (N_i)$ is the $\delta$--function on singular
submanifold $N_i$\cite{jiang,gelfand} with the expression
\[
\delta (N_i)=\int_{N_i}\frac 1{\sqrt{g_x}}\delta ^4(\vec{x}-\vec{z}%
_i(u^1,u^2))\sqrt{g_u}d^2u. 
\]

From the above equation, we conclude that the inner structure of $j^{\mu \nu
}$ or $j^{AB}$ is labelled by the total expansion of $\delta (\vec{\phi})$,
which includes the topological information $\beta _i$ and $\eta _i.$
It is
obvious that, in (\ref{2.5}), when $u^1$ and$\,u^2$ are taken to be time-like
evolution parameter and space-like string parameter, respectively, the inner
structure of $j^{\mu \nu }$ or $j^{AB}$ just represents $l$ strings moving
in the 4--dimensional Riemann-Cartan manifold $X$. The 2-dimensional
singular submanifolds $N_i\,\,(i=1,\cdot \cdot \cdot l)$ are their world
sheets. The Hopf indices $\beta _i$ and Brouwer degree $\eta _i$ classify these
strings. In detail, the Hopf indices $\beta _i$ characterize the absolute
values of the topological quantization and the Brouwer degrees $\eta _i=+1$
correspond to strings while $\eta _i=-1$ to antistrings.

\section{The branch process of strings at limit points}

But from the above discussion, we know that the
results mentioned above are obtained under the condition $J(\phi/v)|_{p_i}\neq 0$.
When this condition fails, i.e. the Brouwer degrees $\eta_i$ are indefinite, what
will happen? In what follows, we will study the case when $J(\phi/v)|_{p_i}= 0$.
It often happens when the zero of $\vec \phi$ includes some branch points, which
lead to the bifurcation of the topological current.

In order to discuss the evolution of these strings and to
simplify our study, we select the parameter $u^1$ as the evolution parameter 
$t$, and let the string parameter $u^2=\sigma $ be fixed. In this case, the
Jacobian matrices are reduced to 
\[
J^{A4}\equiv J^A,\;\;\;J^{AB}=0,\;\;J^3=J^{34}=J(\frac \phi
v),\;\;\;\;A,B=1,2,3, 
\]
for $y^4=u^2\equiv \sigma $. The branch points are determined by 
\begin{equation}
\left\{ 
\begin{array}{l}
\phi ^1(v^1,v^2,t,\sigma )=0 \\ 
\phi ^2(v^1,v^2,t,\sigma )=0 \\ 
\phi ^3(v^1,v^2,t,\sigma )\equiv J(\frac \phi v)=0
\end{array}
\right.  \label{2.88}
\end{equation}
for the fixed $\sigma $, and they are denoted as $(t^{*},p_i)$. In $\phi $%
-mapping theory usually there are two kinds of branch points, namely the
limit points and the bifurcation points\cite{kubicek1}, each kind of them corresponds
to different cases of branch process.

First, in this section, we study the case that the zeros of the order parameter
field $\vec \phi$ includes some limit points which satisfying 
\begin{equation}
J^A(\frac \phi y)|_{(t^{*},p_i)}\neq 0,\;\;\;\;\;A=1 \;\;\;\; or \;\;\;\;2  \label{2.89}
\end{equation}
For simplicity, we consider $A=1$ only.

For the purpose of using the implicit function theorem to study the branch
properties of sstrings at the limit points, we use the Jacobian $J^1(\frac
\phi y)$ instead of $J(\frac \phi v)$ to search for the solutions of $\vec{%
\phi}=0$. This means we have replaced $v^1$ by $t$. For clarity we rewrite
the first two equations of (\ref{2.88}) as
\begin{equation}
\phi ^a(t,v^2,v^1,\sigma )=0,\ \;\;\;\;a=1,2.  \label{2.91}
\end{equation}
Taking account of (\ref{2.89}) and using the implicit function theorem, we
have a unique solution of the equations (\ref{2.91}) in the neighborhood of
the limit point $(t^{*},p_i)$ 
\begin{equation}
t=t(v^1,\sigma ),\ \;\;\;\;v^2=v^2(v^1,\sigma )  \label{2.92}
\end{equation}
with $t^{*}=t(p_i^1,\sigma )$. In order to show the behavior of the strings
at the limit points, we will investigate the Taylor expansion of (\ref{2.92})
in the neighborhood of $(t^{*},p_i)$. In the present case, from (\ref{2.89})
and the last equation of (\ref{2.88}), we get
\[
\frac{dv^1}{dt}=\frac{J^1(\frac \phi y)}{J(\frac \phi v)}|_{(t^{*},p_i)}=%
\infty 
\]
i.e. 
\[
\frac{dt}{dv^1}|_{(t^{*},p_i)}=0. 
\]
Then, the Taylor expansion of $t=t(v^1,\sigma )$ at the limit point $%
(t^{*},p_i)$ is 
\[
t=t(p_i^1,\sigma )+\frac{dt}{dv^1}|_{(t^{*},p_i)}(v^1-p_i^1)+\frac 12\frac{%
d^2t}{(dv^1)^2}|_{(t^{*},p_i)}(v^1-p_i^1)^2 
\]
\[
=t^{*}+\frac 12\frac{d^2t}{(dv^1)^2}|_{(t^{*},p_i)}(v^1-p_i^1)^2. 
\]
Therefore 
\begin{equation}
t-t^{*}=\frac 12\frac{d^2t}{(dv^1)^2}|_{(t^{*},p_i)}(v^1-p_i^1)^2  \label{2.93}
\end{equation}
which is a parabola in $v^1$---$t$ plane. From (\ref{2.93}) we can obtain two
solutions $v_{(1)}^1(t,\sigma )$ and $v_{(2)}^1(t,\sigma )$, which give the
branch solutions of strings at the limit points. If $\frac{d^2t}{(dv^1)^2}%
|_{(t^{*},z_i)}>0$, we have the branch solutions for $t>t^{*}$ (Fig 1(a)), otherwise,
we have the branch solutions for $t<t^{*}$ (Fig 1(b)). Since the topological current of
strings is identically conserved, the topological quantum numbers of these
two generated strings must be opposite at the limit point, i.e. $\beta _1\eta _1+\beta _2%
\eta _2=0$, the former is related to the
creation of cosmic strings and antistrings in pair at the limit points, and the latter
to the annihilation of the cosmic strings.

\section{The Branch process of strings at bifurcation points}

In the following, let us turn to consider the case of bifurcation point in
which the additional restrictions are 
\begin{equation}  \label{2.94}
J^1(\frac \phi y)|_{(t^{*},p_i)}=0,\ \;\;\;\;J^2(\frac \phi
y)|_{(t^{*},p_i)}=0.
\end{equation}
These two restrictive conditions will lead to an important fact that the
function relationship between $t$ and $v^1$ or $v^2$ is not unique in the
neighborhood of bifurcation point $(t^{*},p_i)$. The equation 
\begin{equation}  \label{2.95}
\frac{dv^1}{dt}=\frac{J^1(\frac \phi y)}{J(\frac \phi v)}|_{(t^{*},p_i)}
\end{equation}
which under restraint of (\ref{2.94}) directly shows that the direction of the
integral curve of (\ref{2.95}) is indefinite at the point $(t^{*},p_i)$. This
is why the very point $(t^{*},p_i)$ is called a bifurcation point of the
multistring current. With the aim of finding the different directions of all
branch curves at the bifurcation point, we suppose that 
\begin{equation}  \label{2.96}
\frac{\partial \phi ^1}{\partial v^2}|_{(t^{*},p_i)}\neq 0.
\end{equation}
From $\phi ^1(v^1,v^2,t,\sigma )=0$, the implicit function theorem says that
there exists one and only one function relationship 
\begin{equation}  \label{2.97}
v^2=v^2(v^1,t,\sigma )
\end{equation}
with the partial derivatives $f_1^2=\partial v^2/\partial v^1$, $%
f_t^2=\partial v^2/\partial t$. Substituting (\ref{2.97}) into $\phi ^1$, we
have 
\[
\phi ^1(v^1,u^2(v^1,t,\sigma ),t,\sigma )\equiv 0 
\]
which gives 
\begin{equation}  \label{2.98}
\frac{\partial \phi ^1}{\partial v^2}f_1^2=-\frac{\partial \phi ^1}{\partial
v^1},\ \;\;\;\;\;\;\frac{\partial \phi ^1}{\partial v^2}f_t^2=-\frac{%
\partial \phi ^1}{\partial t},
\end{equation}
\[
\frac{\partial \phi ^1}{\partial v^2}f_{11}^2=-2\frac{\partial ^2\phi ^1}{%
\partial v^2\partial v^1}f_1^2-\frac{\partial ^2\phi ^1}{(\partial v^2)^2}%
(f_1^2)^2-\frac{\partial ^2\phi ^1}{(\partial v^1)^2}, 
\]
\[
\frac{\partial \phi ^1}{\partial v^2}f_{1t}^2=-\frac{\partial ^2\phi ^1}{%
\partial v^2\partial t}f_1^2-\frac{\partial ^2\phi ^1}{\partial v^2\partial
v^1}f_t^2-\frac{\partial ^2\phi ^1}{(\partial v^2)^2}f_t^2f_1^2-\frac{%
\partial ^2\phi ^1}{\partial v^1\partial t}, 
\]
\[
\frac{\partial \phi ^1}{\partial v^2}f_{tt}^2=-2\frac{\partial ^2\phi ^1}{%
\partial v^2\partial t}f_t^2-\frac{\partial ^2\phi ^1}{(\partial v^2)^2}%
(f_t^2)^2-\frac{\partial ^2\phi ^1}{\partial t^2}, 
\]
where 
\[
f_{11}^2=\frac{\partial ^2v^2}{(\partial v^1)^2},\ \;\;\;f_{1t}^2=\frac{%
\partial ^2v^2}{\partial v^1\partial t},\ \;\;\;f_{tt}^2=\frac{\partial ^2v^2%
}{\partial t^2}. 
\]
From these expressions we can calculate the values of $%
f_1^2,f_t^2,f_{11}^2,f_{1t}^2$ and $f_{tt}^2$ at $(t^{*},p_i)$.

In order to explore the behavior of the string at the bifurcation points,
let us investigate the Taylor expansion of 
\begin{equation}  \label{2.99}
F(v^1,t,\sigma )=\phi ^2(v^1,v^2(v^1,t,\sigma ),t,\sigma )
\end{equation}
in the neighborhood of $(t^{*},p_i)$, which according to the Eqs.(\ref{2.88})
must vanish at the bifurcation point, i.e. 
\begin{equation}  \label{2.100}
F(t^{*},p_i)=0.
\end{equation}
From (\ref{2.99}), the first order partial derivatives of $F(v^1,t,\sigma )$
with respect to $v^1$ and $t$ can be expressed by 
\begin{equation}  \label{2.101}
\frac{\partial F}{\partial v^1}=\frac{\partial \phi ^2}{\partial v^1}+\frac{%
\partial \phi ^2}{\partial v^2}f_1^2,\ \;\;\;\frac{\partial F}{\partial t}=%
\frac{\partial \phi ^2}{\partial t}+\frac{\partial \phi ^2}{\partial v^2}%
f_t^2.
\end{equation}
Making use of (\ref{2.98}), (\ref{2.101}) and Cramer's rule, it is easy to prove
that the two restrictive conditions (\ref{2.94}) can be rewritten as
\[
J(\frac \phi v)|_{(t^{*},p_i)}=(\frac{\partial F}{\partial v^1}\frac{%
\partial \phi ^1}{\partial v^2})|_{(t^{*},p_i)}=0, 
\]
\[
J^1(\frac \phi y)|_{(t^{*},p_i)}=(\frac{\partial F}{\partial t}\frac{%
\partial \phi ^1}{\partial v^2})|_{(t^{*},p_i)}=0, 
\]
which give 
\begin{equation}  \label{2.102}
\frac{\partial F}{\partial v^1}|_{(t^{*},p_i)}=0,\ \;\;\;\frac{\partial F}{%
\partial t}|_{(t^{*},p_i)}=0
\end{equation}
by considering (\ref{2.96}). The second order partial derivatives of the
function $F$ are easily to find out to be 
\[
\frac{\partial ^2F}{(\partial v^1)^2}=\frac{\partial ^2\phi ^2}{(\partial
v^1)^2}+2\frac{\partial ^2\phi ^2}{\partial v^2\partial v^1}f_1^2+\frac{%
\partial \phi ^2}{\partial v^2}f_{11}^2+\frac{\partial ^2\phi ^2}{(\partial
v^2)^2}(f_1^2)^2 
\]
\[
\frac{\partial ^2F}{\partial v^1\partial t}=\frac{\partial ^2\phi ^2}{%
\partial v^1\partial t}+\frac{\partial ^2\phi ^2}{\partial v^2\partial v^1}%
f_t^2+\frac{\partial ^2\phi ^2}{\partial v^2\partial t}f_1^2+\frac{\partial
\phi ^2}{\partial v^2}f_{1t}^2+\frac{\partial ^2\phi ^2}{(\partial v^2)^2}%
f_1^2f_t^2 
\]
\[
\frac{\partial ^2F}{\partial t^2}=\frac{\partial ^2\phi ^2}{\partial t^2}+2%
\frac{\partial ^2\phi ^2}{\partial v^2\partial t}f_t^2+\frac{\partial \phi ^2%
}{\partial v^2}f_{tt}^2+\frac{\partial ^2\phi ^2}{(\partial v^2)^2}(f_t^2)^2 
\]
which at $(t^{*},p_i)$ are denoted by 
\begin{equation}  \label{2.103}
A=\frac{\partial ^2F}{(\partial v^1)^2}|_{(t^{*},p_i)},\ \;\;B=\frac{%
\partial ^2F}{\partial v^1\partial t}|_{(t^{*},p_i)},\ \;\;C=\frac{\partial
^2F}{\partial t^2}|_{(t^{*},p_i)}.
\end{equation}
Then, taking notice of (\ref{2.100}), (\ref{2.102}) and (\ref{2.103}), we can
obtain the Taylor expansion of $F(v^1,t,\sigma )$ in the neighborhood of the
bifurcation point $(t^{*},p_i)$ 
\[
F(v^1,t,\sigma )=\frac 12A(v^1-p_i^1)^2+B(v^1-p_i^1)(t-t^{*})+\frac
12C(t-t^{*})^2 
\]
which by (\ref{2.99}) is the behavior of $\phi ^2$ in this region. Because of
the second equation of (\ref{2.88}), we get
\[
A(v^1-p_i^1)^2+2B(v^1-p_i^1)(t-t^{*})+C(t-t^{*})^2=0 
\]
which leads to 
\begin{equation}  \label{2.104}
A(\frac{dv^1}{dt})^2+2B\frac{dv^1}{dt}+C=0
\end{equation}
and 
\begin{equation}  \label{2.105}
C(\frac{dt}{dv^1})^2+2B\frac{dt}{dv^1}+A=0.
\end{equation}
The different directions of the branch curves at the bifurcation point are
determined by (\ref{2.104}) or (\ref{2.105}). There are four possible cases:

\begin{itemize}
\item[Case1]  $(A\neq 0)$: For $\Delta =B^2-AC>0$, from (\ref{2.104}) we
get two different solutions 
\begin{equation}
\frac{dv^1}{dt}\mid _{1,2}=\frac{-B\pm \sqrt{B^2-AC}}A,  \label{2.bifa40}
\end{equation}
which is shown in Fig. 2, where two
cosmic strings collide at the bifurcation point $(t^{*},p_i)$.
This shows that two cosmic strings meet and then depart at the
bifurcation point.
\end{itemize}

\begin{itemize}
\item[Case2]  $(A\neq 0)$: For $\Delta =B^2-AC=0$, there is only one solution
\begin{equation}
\frac{dv^1}{dt}=-B /A,
\end{equation}
which includes three important cases shown in Fig. 3. Firstly, two cosmic strings
tangentially collide at the bifurcation point (Fig. 3(a)). Secondly, two cosmic
strings merge into one cosmic string at the bifurcation point (Fig.
3(b)). Thirdly, one cosmic string splits into two cosmic strings at the
bifurcation point (Fig. 3(c)).
\end{itemize}

\begin{itemize}
\item[Case3]  $(A =0,\,C \neq 0$): For $\Delta =B^2-AC>0$, from (\ref{2.105}) we
have 
\begin{equation}
\frac{dt}{dv^1}\mid _{1,2}=\frac{-B\pm \sqrt{B^2-AC}}C=\left\{ 
\begin{array}{c}
0 \\ 
-\frac{2B}C.
\end{array}
\right.  \label{2.bifa41}
\end{equation}
As shown in Fig. 4, there are two important cases:
(a)One cosmic string splits into three cosmic strings at the
bifurcation point (Fig. 4(a)). (b) Three cosmic strings merge into one at
the bifurcation point (Fig. 4(b)).
\end{itemize}

\begin{itemize}
\item[Case4]  $(A=C=0)$: The equations (\ref{2.104}) and (\ref{2.105}) give
respectively: 
\begin{equation}
\frac{dv^1}{dt}=0,\ \quad \quad \frac{dt}{dv^1}=0.  \label{2.bifa42}
\end{equation}
This case is obvious as in Fig. 5, which is similar to the third situation.
\end{itemize}

The remainder component $dv^2/dt$ can be given by 
\[
\frac{dv^2}{dt}=f_1^2\frac{dv^1}{dt}+f_t^2 
\]
where partial derivative coefficients $f_1^2$ and $f_t^2$ have been
calculated in (\ref{2.98}).

At the end of this section, we conclude that in our string theory there
exist the crucial case of branch process. This means that, when an original
string moves through the bifurcation point in the early universe, it may
split into two strings moving along different branch curves. Since the
topological current of strings is identically conserved, the sum of the
topological quantum numbers of these two splitted strings must be equal to
that of the original string at the bifurcation point, i.e. 
\[
\beta _{i_1}\eta _{i_1}+\beta _{i_2}\eta _{i_2}=\beta _i\eta _i 
\]
for fixed $i$. This can be looked upon as the topological reason of string
splitting.

\section{Conclusion}

In this paper, with the gauge potential decomposition and the so called $%
\phi $-mapping method, we obtain the topological current to describe the
strings in 4--dimensional Riemann--Cartan manifold. In the early universe,
by discussing the properties of the zero points of the vector field $\phi $
and the expansion of the delta function $\delta (\vec{\phi})$, we get the
topological quantization of the strings under the condition that the
Jacobian $J(\frac \phi v)\neq 0$, and pointed out that the singular
manifolds are just the evolution manifolds of these strings. When the Jacobian 
$J(\frac \phi v)=0$, i.e. at the critical points of $\phi$-mapping,
 it is shown that there exist the crucial case of branch
process. Based on the implicit function theorem and the Taylor expansion,
the origin and bifurcation of the strings are detailed in the neighborhoods
of the limit points and bifurcation points of $\phi $-mapping respectively,
i.e. the branch solutions at the limit points and the different directions
of all branch curves at the bifurcation points are calculated out. Because
the topological current of these strings is identically conserved, the
topological charges of these strings will keep to be constant during the
branch processes, which means that the
topological quantum numbers of the two generated string currents must be
opposite at the limit point and, at the bifurcation point, the sum of the
topological quantum numbers of the splitted strings must be equal to
that of the original.

\section*{Figures' Captions}

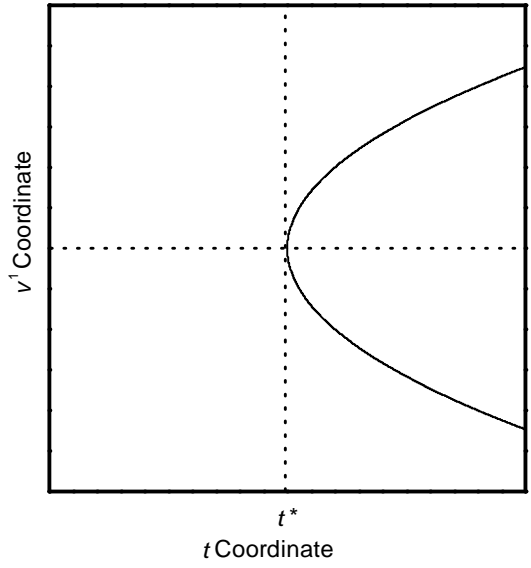
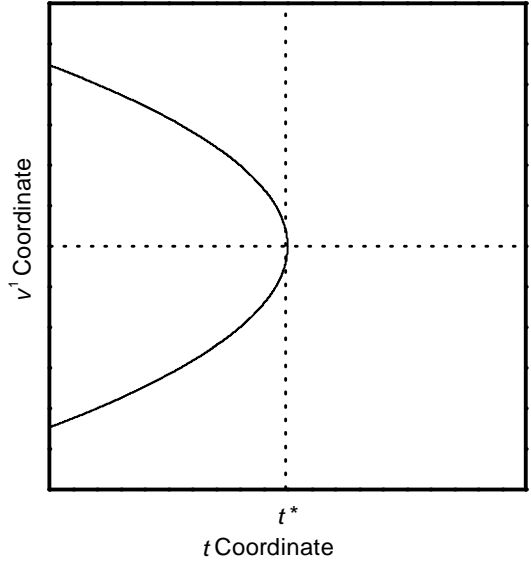
Fig. 1. (a) The creation of two cosmic strings. (b) Two cosmic strings
annihilate in collision at the limit point.

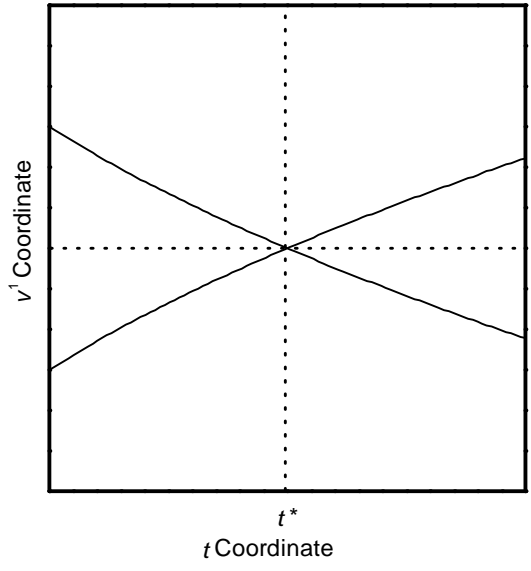
Fig. 2. Two cosmic strings collide with different directions of motion at
the bifurcation point.

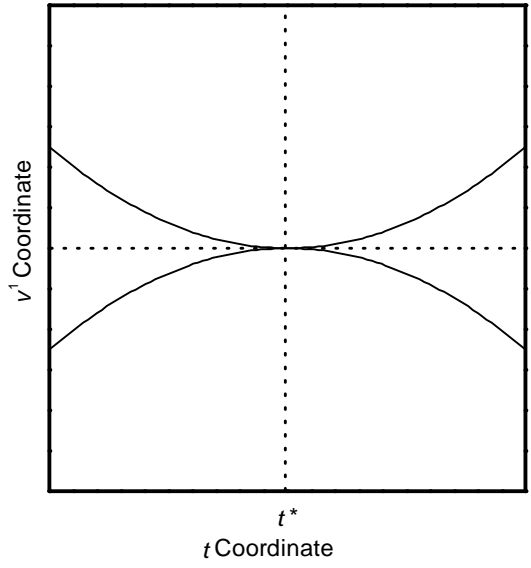
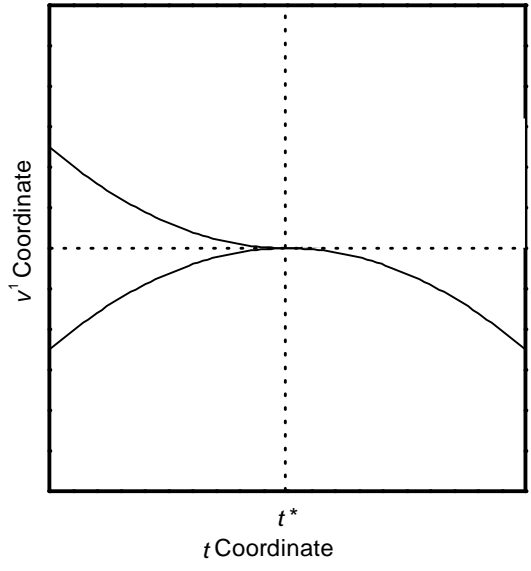
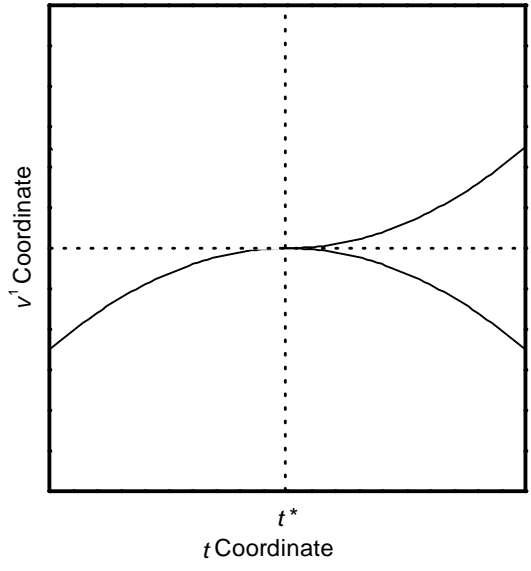
Fig. 3. Cosmic strings have the same direction of motion. (a) Two cosmic strings
tangentially collide at the bifurcation point. (b) Two cosmic strings
merge into one cosmic string at the bifurcation point. (c)
One cosmic string splits into two cosmic strings at the bifurcation
point.

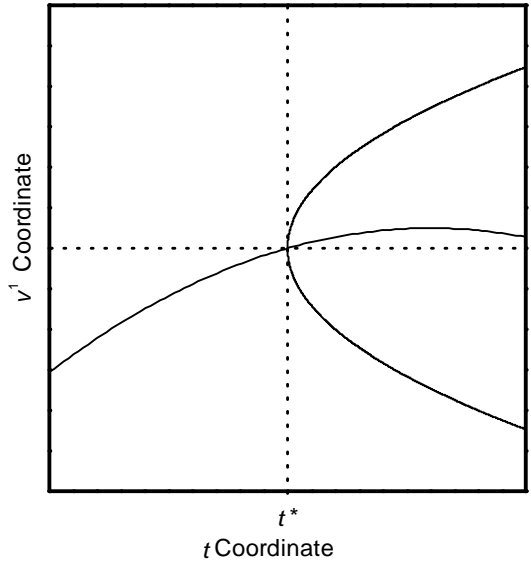
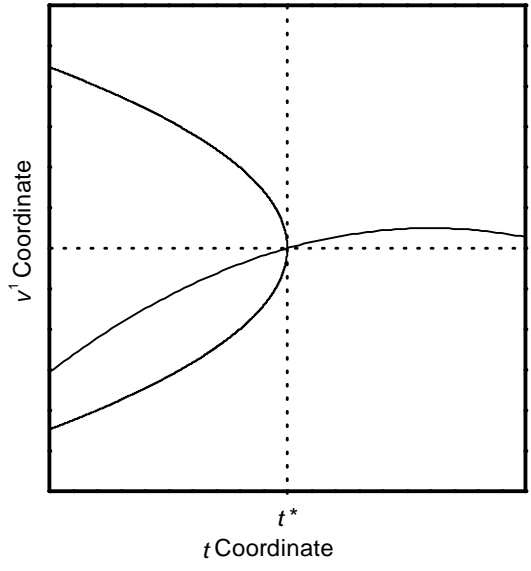
Fig. 4. (a) One cosmic string splits into three cosmic strings at the
bifurcation point. (b) Three cosmic strings merge into one cosmic string
at the bifurcation point.

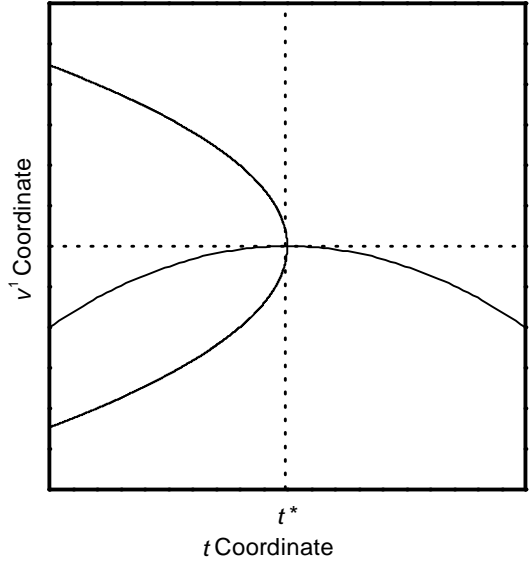
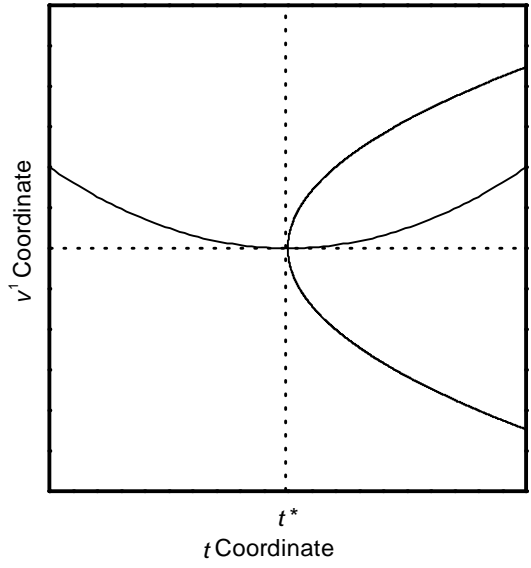
Fig. 5. This case is similar to Fig. 5. (a) Three cosmic strings merge
into one cosmic string at the bifurcation point. (b) One cosmic string
splits into three cosmic strings at the bifurcation point.

\end{document}